\documentclass[twocolumn,aps,prl,showpacs,superscriptaddress]{revtex4}
\usepackage{graphicx}
\usepackage{amsfonts}
\usepackage[varg]{txfonts}
\usepackage{bm}

\begin{document}

\title{Effective masses in a strongly anisotropic Fermi liquid}
\author{Tudor D. Stanescu}
\affiliation{Department of Physics and Joint Quantum Institute,
University of Maryland, College Park, MD 20742-4111},
\author{Victor Galitski}
\affiliation{Department of Physics and Joint Quantum Institute,
University of Maryland, College Park, MD 20742-4111}
\author{H.~D.~Drew}
\affiliation{CNAM, University of Maryland, College Park, MD
20742-4111}

\begin{abstract}
Motivated by the recent experimental observation of quantum
oscillations in the underdoped cuprates, we study the cyclotron and
infrared Hall effective masses in an anisotropic Fermi liquid
characterized by an angle-dependent quasiparticle residue $Z_{\bf
q}$, which models an arc-shaped Fermi surface. Our primary
motivation is to explain the relatively large value of the
cyclotron effective mass observed experimentally and its relation
with the effective Hall mass. In the framework of a
phenomenological model of an anisotropic Fermi liquid, we find
that the cyclotron mass is enhanced by a factor $\left\langle
1/Z_{\bf q} \right\rangle$, while the effective Hall mass
is proportional to $\left\langle Z_{\bf q} \right\rangle /\left\langle
Z_{\bf q}^2 \right\rangle$, where $\langle \cdots \rangle$ implies
an averaging over the Fermi surface. We conclude that if the
$Z$-factor becomes small in some part of the Fermi surface
(e.g., in the case of a Fermi arc), the
cyclotron mass is enhanced sharply while the infrared Hall mass
may remain small. Possible future experiments are discussed.
\end{abstract}

\maketitle

Recently, the observation of magnetic quantum oscillations was
reported in YBCO single crystals
~\cite{QO1_Leyraud,QO2_Bangura,QO3_Yelland,QO4_Jaudet} (ortho-II
YBa$_2$Cu$_3$O$_{6.5}$ and YBa$_2$Cu$_4$O$_8$)  with periods in
$1/H$ indicative of small Fermi surface (FS) pockets (area $\sim
2.4\%$ of the Brillouin zone).  These materials are examples of
hole doped cuprates in the pseudogap region of the phase diagram
where the ubiquitous anomalous physical properties have attracted
much attention. ARPES
measurements~\cite{ARPES1_Damasc,ARPES2_Shen}, for example, have
revealed that portions of the full Fermi surface observed in
optimally doped and overdoped  cuprates are obliterated in the
pseudogap phase, leaving behind so-called Fermi arcs.  However
these ARPES Fermi arcs are not closed surfaces and how they could
figure into the quantum oscillations is the subject of much
current discussion. The magnetic quantum oscillations experiments
probe the geometry of the Fermi surface and the quasiparticle
dynamics in these strongly correlated systems.  The exotic
properties of these materials are widely believed to be tied to
the competition between  strong fluctuations associated with
magnetic, charge and superconducting order, occuring close to a
magnetically ordered Mott state. Interactions with these
fluctuations also govern the quasiparticle dynamics.

One surprising result in the experiments is that the observed
cyclotron frequencies, $\omega_C= eB/(m_C c)$, seem quite small
for the size of the pockets. The cyclotron frequency measured by
the damping of the quantum oscillations is the inverse time for
the quasiparticles to traverse a closed complete orbit around the
Fermi surface. The observed cyclotron masses, $m_C$, are
comparable with the quasiparticle masses estimated for the full
unreconstructed FS (area $\sim 70\%$ of the Brillouin zone) of
optimally doped cuprates obtained from ARPES and other
experiments. On very general grounds $m_C$ is expected to scale
with the pocket size as $m_C\sim \Delta k/v_F$,  where the Fermi
pocket area is $\sim \Delta k^2$ and $v_F$ is the Fermi velocity
which can be measured directly by ARPES. Therefore, developing an
understanding of the value of $m_C$ observed in experiment is
important to understanding the interactions in the pseudogap
phase.

Other experimental probes measure characteristic masses which
depend differently on the k-dependent dynamical mass tensor on the
FS.  ARPES measures the k-dependent quasiparticle dispersion from
which $m_k$ can be estimated. Specific heat, optics and magneto
transport measurements also produce masses which correspond to
different averages of the k-dependent mass on and near the Fermi
surface.  In particular the Hall Effect performed at infrared (IR)
frequencies~\cite{IRH1_Dagan,IRH2_Peng,IRH3_Rigal,IRH4_Shi,IRH5_Zimm}
is characterized by a Hall mass, $m_H$.  Experiments measuring
this quantity in several hole doped cuprates have revealed that
$m_H$, which is comparable to the ARPES masses in optimally doped
materials, decreases rapidly in underdoped cuprates as expected
from the above argument.  It is surprising, however, that the
values of the Hall and cyclotron masses appear to violate the
relation $m_H \ge m_C$, which occurs for models of simple convex
Fermi surfaces for weakly-interacting quasiparticles. For example,
if $m_1$ and $m_2$ are the components of the mass tensor
characterizing an elliptical FS, $m_C = \sqrt{m_1m_2}$ can be
significantly smaller than $m_H = (m_1+m_2)/2$ for an elongated
pocket.

In this paper we propose an explanation for the behavior of the
effective masses based on a model of a highly renormalized Fermi
liquid characterized by a reconstructed FS and strongly
momentum-dependent quasiparticle properties. The existence of a
reconstructed FS in the underdoped cuprates is suggested by the
observation of quantum oscillations, as well as by ARPES and IR
Hall measurements.  Wether this reconstructed FS consists of
closed Fermi pockets or disconnected Fermi arcs is still a matter
of debate. In this work we assume that the first possibility is
realized and we study the implications for the behavior of the
effective masses. One scenario that could lead to a FS
reconstruction involves some type of order, such as
antiferromagnetism~\cite{ORD_Lin}, d-density
wave~\cite{ORD_Chakra}, or stripe\cite{ORD_Millis} order, present
in the system. Another possibility is that the reconstruction is
due to the strong electron correlations generated by the proximity
to a Mott phase~\cite{DMFT1_SK,DMFT2_SCP,DMFT3_Civelli}. In this
scenario the formation of small Fermi pockets is the result of
strong finite range correlations and no broken symmetry needs to
be invoked.  Moreover this scenario does not require energy gaps
in the quasiparticle dispersion relations, in contrast to the
broken symmetry phases, which is consistent with the experimental
observations of optics, IR Hall and ARPES measurements. While this
scenario does not exclude the existence of broken symmetry phases
at low temperatures~\cite{DMFT3_Civelli}, it emphasizes the key
role of competing short-range correlations in the physics of the
underlying ``normal'' state. At a formal level, the FS
reconstruction is the consequence of a large, strongly
momentum-dependent self-energy~\cite{DMFT1_SK}. The self-energy
effects lead to a k-dependent low-energy physics having two major
characteristics: 1) Fermi liquid-like quasiparticles exist along a
reconstructed FS that shrinks as doping is reduced, while a
(pseudo)-gap opens in other regions of the Brillouin zone. 2) The
properties of the quasiparticles along the reconstructed FS are
highly anisotropic. Consequently, physical quantities describing
the quasiparticle are momentum-dependent and contribute to
experimentally measured quantities through specific averages over
the FS. It is therefore natural to address the following
questions: What is the experimental signature of a highly
anisotropic Fermi liquid? In particular, what is the effect of the
anisotropy on the effective masses extracted from various
measurements?

\begin{figure}[tbp]
\begin{center}
\includegraphics[width=0.45\textwidth]{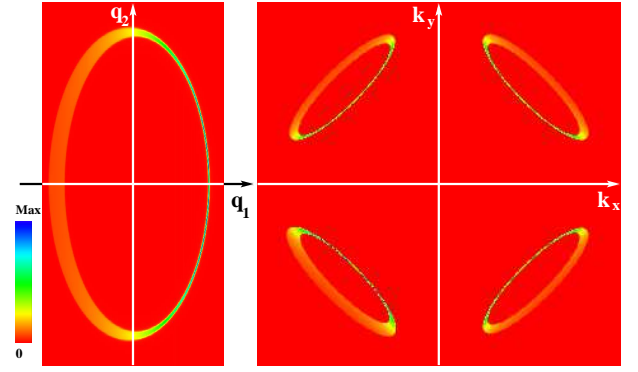}
\end{center}
\caption{(Color online):~ Spectral weight at the Fermi energy for
a Fermi pocket described by the Eq. (\ref{Gq}) with $m_2=4 m_1$
and $\alpha_{\bf q} = 5(q_1/q_{F}^{(1)} -1)^2$ (left). The
spectral weight is obtained by integrating $A(\omega, {\bf q}) =
-{\rm Im}G(\omega, {\bf q})/\pi$ over a small frequency window
about the Fermi energy. The right panel shows a symmetric
distribution of Fermi pockets in momentum space. }  \vspace*{-0.15in} \label{Fig1}
\end{figure}

To address these questions we study a phenomenological model of an
anisotropic Fermi liquid, which mimics the arc-shaped Fermi
pockets and may lead to the experimentally observed quantum
oscillations. We are not concerned here with the physical
mechanism responsible for the existence of these pockets. Our main
assumption is that the low-energy physics is determined by
quasiparticles with highly anisotropic properties. Moreover, we
focus on the effects of an anisotropic self-energy that produces a
strongly angle-dependent quasiparticle residue $Z_{\bf q}$. Our
starting point is the Fermi-liquid Green function
\begin{equation}
G(\omega, {\bf q}) = \frac{1}{\omega(1+\alpha_{\bf q}) - \xi_{\bf
q} + i\eta}                      \label{Gq}
\end{equation}
where $\xi_{\bf q} = (q_1^2/2m_1 + q_2^2/2m_2) -\epsilon_F$, with
$\epsilon_F$ being the Fermi energy) represents a parabolic bare
band characterized by a two-dimensional elliptic Fermi surface.
The anisotropy along the Fermi surface is determined by the
momentum-dependent self-energy $\Sigma(\omega, {\bf q}) =
-\omega\alpha_{\bf q} -i\eta$. A particular choice of the
anisotropy parameter $\alpha_{\bf q}$ is discussed bellow. Note
that the quasiparticle residue $Z_{\bf q} = (1+\alpha_{\bf
q})^{-1}$ represents the spectral weight of the quasiparticle peak
in the spectral function. Consequently, in the presence of a large
anisotropy, the low-energy spectral weight associated with the
Fermi surface will have an arc-like shape (see Fig. \ref{Fig1}).
We assume that the $\omega^2$ momentum-dependent contribution to
the imaginary part of the self-energy is negligible in the
relevant frequency range. In addition, we make the assumption that
frequency-independent contribution $(-i\eta)$ is weakly
momentum-dependent and we consider it a constant. In general, the
effects of the quasiparticle residue anisotropy, $Z_{\bf q} =
(1+\alpha_{\bf q})^{-1}$, and those determined by the scattering
time anisotropy, $1/2\tau_{\bf q} = \eta_{\bf q}$ are entangled,
but we will operate under these simplifying conditions in order to
clearly identify the relevant effects. Note that Eq. (\ref{Gq})
describes one Fermi pocket in a properly chosen coordinate system.
When calculating  physical observables, we need to restore the
symmetry by averaging over several
 pockets symmetrically arranged in momentum space (see Fig.
\ref{Fig1}). 

We now consider a two-dimensional system described by Eq.
(\ref{Gq}) in the presence of a uniform magnetic field ${\mathbf
B}$ 
oriented along the direction perpendicular to the two-dimensional plane.
The effects of the magnetic field are introduced,
based on the work by Luttinger~\cite{MAG_Lutt} and
Roth~\cite{MAG_Roth}, through the standard substitution ${\mathbf q}
\rightarrow -i{\mathbf \nabla} -\frac{e}{c}{\mathbf A}$ made
both in the bare band $\xi({\bf q})$
and the self-energy~\cite{MAG_Kita}. Consequently, the
anisotropy of the quasiparticle residue has a direct impact on the
effective Landau level structure. In the presence of the
magnetic field, the system is described by the operator
$\hat{G}_{\omega} = [\omega - \hat{K}_{\omega} +i\eta]^{-1}$,
where $\hat{K}_{\omega} = \xi(-i{\mathbf \nabla}
-\frac{e}{c}{\mathbf A}) - \omega~\alpha(-i{\mathbf \nabla}
-\frac{e}{c}{\mathbf A})$. The next step is to find the
eigenvalues of the operator $\hat{K}_{\omega}$ for each frequency.
In order to make this step mathematically tractable, we make the following
explicit choice for the momentum dependence of the anisotropy factor (consistent with
the arc-shaped form of a pocket)
$\alpha_{\bf q} = \alpha \left(\frac{q_1}{q_{F}^{(1)}} -1\right)^2,$
where $\alpha$ is a positive number and $q_{F}^{(1)}
=\sqrt{2m_1\epsilon_F}$ is the Fermi k-vector along the direction
``1''. With this choice, the quasiparticle strength is
continuously reduced from $Z_{(q_F^{(1)}, 0)} = 1$ on one side of
the Fermi pocket to $Z_{(-q_F^{(1)}, 0)} = 1/(1+4\alpha)$ on the
opposite side. Writing the vector potential in the symmetric gauge
as ${\mathbf A} = \left(-\frac{B}{2} x_2, \frac{B}{2} x_1\right)$,
the operator $\hat{K}_{\omega}$ becomes
\begin{eqnarray}
\hat{K}_{\omega} &=& \frac{\left(-i\partial_{x_1} + \frac{e B}{2
c}x_2\right)^2}{2 m_1} + \frac{\left(-i\partial_{x_2} - \frac{e
B}{2 c}x_1\right)^2}{2 m_2}  \nonumber \\
&-&\frac{\alpha\omega}{(q_F^{(1)})^2}\left(-i\partial_{x_1}
+\frac{eB}{2c}x_2 -q_F^{(1)}\right)^2 -\epsilon_F.  \label{Kw}
\end{eqnarray}
The eigenproblem  for this operator can be reduced to the
standard Landau-level problem by first separating a plane-wave factor
$\exp[-ix_1q_F^{(1)}\alpha\omega/(\epsilon_F-\alpha\omega)]$ in
the eigenfunction, then rescaling the variables
$x_1\rightarrow\tilde{x}_1 \lambda$, $x_2\rightarrow
\tilde{x}_2/\lambda$ with $\lambda =
(m_2/m_1)^{1/4}(1-\alpha\omega/\epsilon_F)^{1/4}$. The
corresponding ``Landau levels'' are
\begin{equation}
W_n(\omega) =
\omega_C\sqrt{1-\frac{\alpha\omega}{\epsilon_F}}\left(n+\frac12\right)
-\frac{\alpha\omega}{1-\frac{\alpha\omega}{\epsilon_F}}
-\epsilon_F,      \label{Wn}
\end{equation}
where $\omega_C = eB/c\sqrt{m_1m_2}$ represents the cyclotron
frequency of the elliptic pocket in the absence of the self-energy
anisotropy. Finally, we identify the physical Landau levels, i.e.,
we determine the frequencies at which the Green function has
poles, by solving the equation $\omega - W_n(\omega) = 0$.
Focusing on the relevant low-frequency regime,
$\alpha\omega\ll\epsilon_F$, we obtain the set of solutions
$\omega_j = \frac{\omega_C}(j + \delta)/({1+\frac32\alpha})$, with
$j=0, \pm1, \pm2, \dots$ and $0\leq \delta< 1$. The effective
cyclotron frequency for our model is $\omega_C^* =
\omega_C/(1+\frac32\alpha)$ and represents the spacing between
Landau levels in the vicinity of Fermi energy. This spacing
determines the temperature-dependent damping of the amplitude of
quantum oscillations through the standard Lifshitz-Kosevich
formula. Consequently, the effective cyclotron mass extracted from
an analysis of the temperature dependence of the amplitude of
quantum oscillations is $m_C^* = \sqrt{m_1m_2}(1+\frac32\alpha)$.
Note that the enhancement factor of the effective mass due to
self-energy effects is exactly $(1+\frac32\alpha) = \left\langle
1/Z_{\bf q}\right\rangle$, where $\left\langle \dots\right\rangle$
represents an average over the Fermi surface, and we have
\begin{equation}
m_C^* = \sqrt{m_1m_2}\left\langle \frac{1}{Z_{\bf
q}}\right\rangle. \label{mc}
\end{equation}
The cyclotron mass is enhanced by a factor equal to the average of
$1/Z_{\bf q}$ over the Fermi surface. Note that, the quantum
oscillations are observable if the quasiparticles have  a finite
weight everywhere along a closed Fermi surface, as expected from
the semi-classical picture. If the $Z_{\bf q}$ anisotropy is
large, the main contribution to the mass comes from the regions
with small quasiparticle residue. We note that, within our model,
a similar enhancement obtains for the effective thermodynamic mass
extracted from the specific heat, $m^* \sim C/T$, and we get $m^*
= m_C^*$.

Let us turn now our attention to the evaluation of the
longitudinal and Hall dynamical conductivities for the anisotropic
Fermi liquid described by Eq. (\ref{Gq}). We neglect contributions
coming from the current vertex corrections and focus on a
frequency regime characterized by $\omega_C \ll  \omega$, $1/\tau
\ll \epsilon_F$ with $\omega\tau$ arbitrary, where the cyclotron
frequency $\omega_C$ is the smallest energy scale in the problem
and $1/\tau=\eta$ is the inverse scattering time.  We also assume
a symmetric arrangement of the Fermi pockets (see Fig.
\ref{Fig1}), so the final values for the conductivities represent
averages over the  possible orientations. Using the standard Kubo
formalism we have
\begin{eqnarray}
\sigma_{\alpha\alpha}(\omega) &=& \frac{-i
e^2}{\omega}T\sum_{\epsilon_n}\sum_{\bf q}
\frac{q_{\alpha}^2}{m_{\alpha}^2}G_n({\bf q})G_{n-}({\bf q}),
\label{sigGG} \\
\sigma_{\alpha\beta}(\omega) &=& \frac{\pm
e^2\omega_C}{\omega}T\sum_{\epsilon_n, ~{\bf q}}
\frac{q_{\alpha}}{\sqrt{m_1m_2}}\left[G_n\partial_{\alpha}G_{n-} -
\partial_{\alpha}G_{n}G_{n-} \right], \nonumber
\end{eqnarray}
where we use the notation $G_n\equiv G(i\epsilon_n, {\bf q}) =
[i\epsilon_n(1+\alpha_{\bf q}) - \xi_{\bf q}
+i~\mbox{sign}(\epsilon_n)/(2\tau)]^{-1}$ for the Matsubara Green
function,  $G_{n-} = G(i\epsilon_n-i\omega_m, {\bf q})$,
$\partial_{\alpha} = \partial/\partial_{q_{\alpha}}$ and
$\alpha\neq\beta$. Here, the analytical continuation to real
frequencies, $i\omega_m \rightarrow \omega - i\delta$ is
implicitly assumed. When performing the sum over ${\bf q}$ in
(\ref{sigGG}), it is convenient to rescale the momentum in order
to have a circular Fermi surface. We define $\tilde{q}_1 =
\sqrt{m_2/m_1}~q_1$ and $\tilde{q}_2 = \sqrt{m_1/m_2}~q_2$, as
well as the angle $\varphi_{\tilde{\bf q}}$ between the vector
$\tilde{\bf q}$ and the ``1''-axis. With these notations, the
expression of the longitudinal conductivity becomes
\begin{equation}
\sigma_{xx} = \frac{n
e^2}{\sqrt{m_1m_2}}\left\langle\frac{\sqrt{\frac{m_2}{m_1}}
\cos^2\varphi_{\tilde{\bf q}} +\sqrt{\frac{m_1}{m_2}}
\sin^2\varphi_{\tilde{\bf q}}}{\frac{1}{\tau} -
i\omega(1+\alpha_{\tilde{\bf q}})}\right\rangle, \label{sig_xx}
\end{equation}
where $n$ is the carrier concentration and $\langle\dots\rangle$
represents the average over the (circular) Fermi surface. 
Similarly, for the Hall conductivity we have
\begin{equation}
\sigma_{xy} = \frac{ne^2}{\sqrt{m_1m_2}}~\omega_C
\left\langle\frac{1}{\left[\frac{1}{\tau} -
i\omega(1+\alpha_{\tilde{\bf q}})\right]^2}\right\rangle,
\label{sig_xy}
\end{equation}
with $\omega_C = eB/c\sqrt{m_1m_2}$ being the bare cyclotron
frequency.  Note that in Eqns. (\ref{sig_xx}) and (\ref{sig_xy})
no assumption has been made about the momentum dependence of the
anisotropy factor $\alpha_{\bf q}$. Also note that similar
expressions can be written in the more general case of a
momentum-dependent scattering time $\tau_{\bf q}$. In practice the
momentum dependence of $\tau$ and also its frequency dependence
have to be taken into account. One can see from equations
(\ref{sig_xx}) and (\ref{sig_xy}) that any quantity extracted from
the low-frequency dynamical conductivities will contain the
combined effect of anisotropic $Z_{\bf q}$ and $\tau_{\bf q}$.

\begin{figure}[tbp]
\begin{center}
\includegraphics[width=0.4\textwidth]{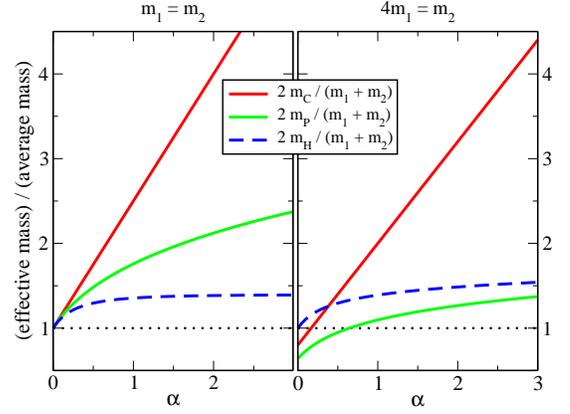}
\end{center}
\caption{(Color online):~ Dependence of the effective masses on
the anisotropy of the Fermi surface. The left and right panels
correspond to circular and elliptical Fermi surfaces,
respectively. The effective cyclotron, plasma and Hall effective
masses are calculated using equations (\ref{mc}),  (\ref{mp}) and
(\ref{mh}). The thermodynamic effective mass extracted from the
specific heat coincides with $m_C$.} \vspace*{-0.15in}
\label{Fig2}
\end{figure}

An interesting and experimentally important regime~\cite{IRH1_Dagan,IRH2_Peng,IRH3_Rigal,IRH4_Shi,IRH5_Zimm} corresponds to infrared frequencies satisfying $\omega\tau \gg 1$. In this case,
$\tau$ becomes irrelevant and the leading contribution to the
longitudinal conductivity takes the standard form $\sigma_{xx}
\approx i\omega_P^2/(4\pi\omega)$, where $\omega_P^2 = 4\pi n
e^2/m_P^*$ is the plasma frequency with
\begin{equation}
m_P^*
=\frac{\sqrt{m_1m_2}}{\left\langle\left(\sqrt{\frac{m_2}{m_1}}\cos^2\varphi_{\tilde{\bf
q}} + \sqrt{\frac{m_1}{m_2}}\sin^2\varphi_{\tilde{\bf
q}}\right)Z_{\tilde{\bf q}} \right\rangle},   \label{mp}
\end{equation}
 Note that the plasma frequency defined here includes only contributions
coming from low-energy quasiparticles in the vicinity of the Fermi
energy. Our phenomenological model does not include incoherent
self-energy contributions that should be present in a strongly
correlated system above a certain energy scale. In discussing the
high-frequency limit we implicitly assume that $\omega$ does not
exceed that energy scale. The leading high-frequency contribution
to the Hall conductivity is $\sigma_{xy} \approx
-\omega_P^2\omega_H /(4\pi \omega^2)$ with $\omega_H = eB/(c
m_H^*)$ being the Hall frequency. The corresponding effective mass
is
\begin{equation}
m_H^* = \frac{m_1m_2}{m_P^*\left\langle Z_{\tilde{\bf
q}}^2\right\rangle}.            \label{mh}
\end{equation}
For $Z_{\bf q} = 1$, Eq. (\ref{mh}) takes the standard form $m_H =
(m_1+m_2)/2$. We note that Eqs.~(\ref{mp}) and
(\ref{mh}) for the effective masses were obtained without making
any assumption about the q-dependence of the anisotropy factor
$\alpha_{\bf q}$ and, implicitly, about the q-dependence of
$Z_{\bf q}$. In addition, as in the case of the cyclotron mass, it
is straightforward to show that the same expressions for the
effective masses can be obtained for a quasiparticle residue of
the form $Z_{\bf q} = Z_0/(1+\alpha_{\bf q})$.

To analyze these results, let us consider, for simplicity, the
case of a circular Fermi pocket. The effective cyclotron mass is
enhanced by a factor $\langle 1/Z_{\bf q}\rangle$, while the
effective plasma mass is increased by $1/\langle Z_{\bf q}\rangle$
and the Hall mass by $\langle Z_{\bf q}\rangle/\langle Z_{\bf
q}^2\rangle$. In the isotropic case, all these enhancement factors
are equal to $1/Z$. However, in the presence of an anisotropy,
$m_C$ is enhanced more strongly, due to the singular nature of
$1/Z_{\bf q}$. In contrast, both $1/\langle Z_{\bf q}\rangle$ and
$1/\langle Z_{\bf q}^2\rangle$ remain relatively small even for
$Z_{\bf q}$ vanishing at certain q-vectors along the Fermi
surface. Consequently, the effective masses extracted from
conductivity measurements will be finite even for a Fermi surface
consisting in disconnected Fermi arcs. The dependence of the
effective masses on the anisotropy factor for $\alpha_{\bf q} =
\alpha \left(q_1 / q_{F}^{(1)} -1\right)^2$ is shown in Fig.
\ref{Fig2}. The left panel corresponds to a circular Fermi pocket,
while the right panel corresponds to the elliptic case. Note that
for an elongated Fermi pocket, in the absence of anisotropy, the
Hall mass is always larger that the cyclotron mass. However, even
a small anisotropy can make $m_C^*$ comparable with or larger than
$m_H^*$.

In conclusion, we have shown that various experimentally relevant
effective masses  represent different averages over the Fermi
surface. The basic ingredients that contribute to the effective
mass are the band structure, which determines the bare mass tensor
(i.e., the values of $m_1$ and $m_2$ in our model), and the
self-energy contributions, which are contained in the
quasiparticle residue $Z_{\bf q}$. It is convenient to regard
$Z_{\bf q}$ as a product of a q-independent factor $Z_0$ and a
momentum-dependent anisotropy factor $1/(1+\alpha_{\bf q})$. The
main message of this letter is that the $Z_{\bf q}$ anisotropy
plays a key role in establishing the value of a given effective
mass. The question concerning the dependence of the anisotropy on
doping is beyond the scope of this study. However, we point out
that the different components contributing to the effective mass
may have a doping evolution with competing effects. For example,
as the doping  is reduced, the pockets are expected to shrink,
leading to a decrease in the bare mass. On the other hand $Z_0$,
the ``nodal'' quasiparticle residue, should decrease when
approaching the Mott insulator, leading to an enhancement of the
effective mass. What about the $Z_{\bf q}$ anisotropy? If pocket
formation is associated with antiferromagnetic
fluctuations~\cite{AFHarrison}, for example, we can speculate that
the anisotropy should be weak at very small doping and increase
strongly as we approach optimal doping. In this case, our analysis
would predict a very sharp increase of the cyclotron mass upon
approaching optimal doping.

A high priority suggested by these considerations is experiments
that include a systematic analysis of the cyclotron and Hall
effective masses and specific heat as a function of doping in a
single material under similar conditions i.e.,
 high magnetic fields and low temperatures.
Comparison of these masses with our theoretical
predictions will help to determine the relation between the
quantum oscillations and the Fermi arcs observed in ARPES and
thereby lead to an understanding of the pseudogap in the
underdoped cuprates.

H. D. Drew was supported by NSF grant DMR-0303112.

\vspace*{-0.1in}
\bibliography{referinte}

\end{document}